\documentstyle[preprint,aps]{revtex}
\let\a=\alpha\let\b=\beta\let\d=\delta
\let\e=\epsilon\let\g=\gamma
\let\l=\lambda
\let\m=\mu\let\n=\nu
\let\t=\tau

\let\D=\Delta\let\F=\Phi

\newcommand{\be}{\begin{equation}}
\newcommand{\ee}{\end{equation}}
\newcommand{\bea}{\begin{eqnarray}}
\newcommand{\eea}{\end{eqnarray}}

\newcommand{\del}{\partial}

\newcommand{\nbox}{{\,\lower0.9pt\vbox{\hrule \hbox{\vrule height 0.2 cm \hskip
0.2 cm \vrule height 0.2 cm}\hrule}\,}}
\begin{document}
\begin{titlepage}
\begin{center}
\vskip .2in
\hfill
\vbox{
    \halign{#\hfil         \cr
           hep-th/0011279 \cr
           RUNHETC-2000-52 \cr
           }  %end of \halign
      }   %end of \vbox
\vskip 0.5cm
{\large \bf Supergravity Duals for N=2 Gauge Theories}\\
\vskip .2in
{\bf Arvind Rajaraman} \footnote{e-mail address:
arvindra@muon.rutgers.edu}
\\
\vskip .25in
{\em
Department of Physics and Astronomy,
Rutgers University,
Piscataway, NJ 08855.  \\}
\vskip 1cm
\end{center}
\begin{abstract}
We construct supergravity solutions for Dp-branes
at orbifold points.
The solutions are written in terms of a single function,
which is the solution to a nonlinear differential equation.
The near horizon limits of these solutions are dual, in the AdS/CFT sense,
to super-Yang-Mills theories with 8 supercharges in various
dimensions.
In particular, we present a dual to
${\cal N}=2$ $SU(N)$ SYM theory in 3+1 dimensions, and
analyse some aspects of the duality.
 \end{abstract}
\end{titlepage}
\newpage

\newpage
\bigskip

\section{Introduction}
The AdS/CFT correspondence\cite{Maldacena:1998re}, which provides
a supergravity dual for ${\cal N}=4$
super-Yang-Mills theories, can be generalized to generate supergravity duals for theories
with
less supersymmetry. In particular, there has been
much interesting recent work on duals to
 ${\cal N}=1$ gauge theories 
(including  ${\cal N}=1$ $SU(N)$ super-Yang-Mills theory)
\cite{Polchinski:2000uf,Klebanov:2000nc,Klebanov:2000hb,Acharya:2000gb,Maldacena:2000yy,MV}.
However,  ${\cal N}=2$ theories have received less
attention.

In this paper, we will construct  supergravity duals
to $SU(N)$ super Yang-Mills theories with 8 supercharges in various dimensions.
This is closely related to the work of \cite{Brinne:2000fh}, where
supergravity solutions corresponding to D4-branes ending on
NS5-branes were found. (This would be a dual to
MQCD, which is in the same
universality class as super-Yang-Mills theory\cite{Witten:1998jd}).

We shall here construct solutions corresponding to branes on orbifolds.
These solutions are complicated; nevertheless, they can be
found by the methods discussed in \cite{Rajaraman:2000ws}. These solutions
are determined by one function, which satisfies
a nonlinear differential equation.
We have not found an explicit solution to
this differential equation, nevertheless, the
solution is determined in principle.

Branes on orbifolds were studied by
Douglas and Moore \cite{Douglas:1996sw} (see also \cite{Johnson:1997py}) who found
a general prescription for the worldvolume theory
of these branes. In particular,
$N$ D5-branes wrapped on a 2-cycle of a $T^4/Z_N$
orbifold have a worldvolume theory which is precisely
4-dimensional $SU(N)$ super Yang-Mills theory with 8 supercharges. 

We can now follow the general reasoning of Maldacena.
We can find the supergravity solution produced
by these D5-branes. The near horizon limit of this geometry
is then dual to the worldvolume theory, i.e. it is
dual to 4-dimensional ${\cal N}=2$ super Yang-Mills theory.

We shall start with a D2-brane on a $T^4/Z_N$
orbifold. It will prove extremely helpful to
consider not the singular limit, but a (partially)
resolved orbifold where one two-cycle has been resolved.
We shall look at a D2-brane wrapped on this resolved 2-cycle.

The
metric for the resolved orbifold is
known exactly to be
\bea
\label{orbmetric}
ds^2=H^{-1}(dx_9+A_7dx^7+A_8dx^8)^2+H(dx_7^2+dx_8^2+dx_6^2)
\eea
with
\bea
\del_6 A_8=\del_7 H ~~~~~~~~~~~~~~~~~~~~~~~~~~~~~\quad \quad \del_6 A_7=-\del_8 H
\nonumber
\\
\del_7 A_8-\del_8 A_7=-\del_6 H~~~~~~~~~~~~~~~~~~~~~~
\nonumber
\eea
(We have chosen an unconventional gauge choice where $A_6=0$.
Retaining $A_6$ leads to a more symmetric set
of equations: see, for example \cite{Johnson:1997py}.)

Here $H$ is a harmonic function satisfying
$(\del_6^2+\del_7^2+\del_8^2)H=0$.
If we want to resolve one 2-cycle in a $Z_N$ orbifold,
we take
\bea
\label{defa}
H={N_1\over x_6^2+x_7^2+x_8^2}+{N_2\over (x_6-a)^2+x_7^2+x_8^2}
\eea

A 2-brane can then extend between the centres at $x_6=0$ and
$x_6=a$, and the second worldvolume direction of the 2-brane
wraps the $x_9$ direction. (There is a periodic identification
of $x_9$ with period $4\pi$.)

The metric for the resolved orbifold is clearly very similar
to the metric of parallel 5-branes. Similarly the 2-brane
wrapping the two cycle is very similar to a brane stretching
between these parallel 5-branes. The supergravity solution can thus
be found using the methods described in \cite{Rajaraman:2000ws} for
the
construction of intersecting brane solutions.
We shall describe this construction in great detail 
in the next section.

Hoever, the orbifold point also has
a nonzero B-field
turned on \cite{Aspinwall:1995zi}. This
B-field induces a D0-brane charge on the D2-brane. We should
therefore look for a solution with both D2-brane and D0-brane charge.
This can be done by lifting the D2-brane solution to
11 dimensions, and boosting it, thereby adding D0-brane charge.
By an appropriate choice of the boost parameter, we
can tune the B-field on the 2-cycle to any value
desired.

The solutions for the other branes on the orbifold can be
found by T-duality. In particular, T-dualizing thrice, 
we get a D5-brane wrapped on the orbifold 2-cycle, which
as we have seen, is dual to 3+1 dimensional
${\cal N}=2$ SYM.

The solution we find is different from other attempts
to realize ${\cal N}=2$ theories\cite{Polchinski:2000mx}. The difference is that in
other cases, the full theory is not exactly $SU(N)$ ${\cal N}=2$ super-Yang-Mills
theory in 3+1 dimensions. In some cases, the theory is regulated
in the UV by a 6-dimensional CFT, and in other cases, it is
regulated by a theory with a larger gauge group in 4-dimensions.
Furthermore, some of these theories live on $S^3\times R$ rather
than $R^{3,1}$.
The constuction we present is dual to the exact
${\cal N}=2$ super-Yang-Mills theory on $R^{3,1}$ with
gauge group $SU(N)$. This is why it differs from
the other constructions.

\section{2-brane on $Z_N$ orbifold}
\subsection{Notation}
We will start by constructing the supergravity solution for
a D2-brane wrapped on a 2-cycle of a $Z_N$ orbifold.
We can approach the SUGRA solution of this
system in the same way as \cite{Rajaraman:2000ws}.
First we establish some notation.

We will denote the directions $x_1,x_2,x_3,x_4,x_5$ collectively
by $x_a$. The directions $x_7,x_8$ will collectively be
called $x_\a$.

The approach in \cite{Rajaraman:2000ws} used an analysis of the condition for
conserved supercharges
\bea
\label{susy}
\del_\mu \e +{1\over 2}\omega_\mu^{\ ab}\g_{ab}\e
+(c_1e^\Phi F_\mu^{\ abc}\g_{abc}+c_2e^\Phi \F_{abcd}g_{\mu}^{\ abcd})\e=0
\eea
where $c_1, c_2$ are constants with $c_1=-c_2$.

We will make the following ansatz (see \cite{Rajaraman:2000ws} for a more
detailed discussion)

\begin{itemize}

\item $\e=(g_{00})^{1\over 4}\e_0$ where $\e_0$ is a constant spinor.

\item The constant spinor satisfies
\bea
\label{projs}
(1-\g^{069})\e_0= (1-\g^{6789})\e_0=0
\eea

\item
The metric components will be taken to be diagonal 
with the addition of $e_{9\tilde{7}}, e_{9\tilde{8}}$
(present  already in (\ref{orbmetric})) and $e_{6\tilde{a}}$ (induced
by the 2brane).

\item
The nonzero gauge field strengths are
\bea
G_{069a}\quad G_{069\a}\quad G_{09a\a}
%\\
\quad H_{078a}\quad H_{0678}\quad H_{06a\a}
\nonumber
\eea

\end{itemize}

As in \cite{Rajaraman:2000ws}, we have chosen to denote the same
field
strength by different letters depending on the indices.
Both $G$ and $H$ are the field strength coupling to 
$D2$ branes.

\subsection{Supersymmetry equations}
The SUSY equations (\ref{susy}) now reduce to a set of
algebraic equations which determine the field
strengths in terms of the spin connections and
also impose
some constraints on the spin connections.

To figure out these algebraic equations, note that the
SUSY constraints (\ref{projs}) are singlets under 
rotations in $x_a, x_\a$. Hence we can decompose the equations
in representations of the rotations in $x_a, x_\a$.

For instance, the field strengths and spin connections
which transform under rotations in $x_1$, but are singlets
under other rotations, are
%\bea
$\omega_A^{\ A1}, G_{0691}$, and $ H_{0781}$.
%\eea
Hence these terms must cancel against each other
in each SUSY equation (\ref{susy}) with any $\mu$.
For $\mu=0,4,6,7,9$, we find respectively
\bea
\omega_0^{\ 01}\g_1+ e^\Phi G_{0691}\g^{0691}+e^\Phi H_{0781}\g^{0781}=0
\nonumber
\\
\omega_4^{\ 41}\g_1- e^\Phi G_{0691}\g^{0691}-e^\Phi H_{0781}\g^{0781}=0
\nonumber
\\
\omega_6^{\ 61}\g_1+ e^\Phi G_{0691}\g^{0691}-e^\Phi H_{0781}\g^{0781}=0
\\
\omega_7^{\ 71}\g_1- e^\Phi G_{0691}\g^{0691}+e^\Phi H_{0781}\g^{0781}=0
\nonumber
\\
\omega_9^{\ 91}\g_1+ e^\Phi G_{0691}\g^{0691}-e^\Phi H_{0781}\g^{0781}=0
\nonumber
\eea
We have chosen a convenient normalization where
$c_1$ is absorbed into the field strengths.

From these equations we find 
\bea
\label{fld1}
e^\Phi G_{0691}\g^{0691}=-{1\over 2}(\omega_0^{\ 01}-\omega_7^{\ 71})\g_1
\\
e^\Phi H_{0781}\g^{0781}=-{1\over 2}(\omega_0^{\ 01}+\omega_7^{\ 71})\g_1
\eea
and the constraints
\bea
\label{cons1}
\omega_4^{\ 41}+\omega_0^{\ 01}=0
\\
\omega_6^{\ 61}+\omega_7^{\ 71}=0
\\
\omega_9^{\ 91}+\omega_7^{\ 71}=0
\eea

We can similarly write down the remaining algebraic 
equations. 
\bea
\g^{0971}G_{0971}+\g^{0681}H_{0681}=0
\nonumber
\\
\omega_1^{\ 67}\g^1_{\ 67}+\omega_1^{\ 89}\g^1_{\ 89}+e^\Phi G_{0971}\g^{0971}+e^\Phi H_{0681}\g^{0681}=0
\nonumber
\\
\omega_6^{\ 17}\g^6_{\ 17}-e^\Phi G_{0971}\g^{0971}+e^\Phi H_{0681}\g^{0681}=0
\nonumber
\\
\omega_7^{\ 16}\g^7_{\ 16}+e^\Phi G_{0971}\g^{0971}-e^\Phi H_{0681}\g^{0681}=0
\nonumber
\\
\omega_8^{\ 19}\g^8_{\ 19}-e^\Phi G_{0971}\g^{0971}+e^\Phi H_{0681}\g^{0681}=0
\nonumber
\\
\omega_9^{\ 18}\g^9_{\ 18}+e^\Phi G_{0971}\g^{0971}-e^\Phi H_{0681}\g^{0681}=0
\nonumber
\\
\omega_0^{\ 07}\g_7+e^\Phi G_{0697}\g^{0697}=0
\nonumber
\\
\omega_a^{\ a7}\g_7-e^\Phi G_{0697}\g^{0697}=0
\nonumber
\\
\omega_6^{\ 67}\g_7+\omega_6^{\ 98}\g^6_{\ 98}+e^\Phi G_{0697}\g^{0697}=0
\nonumber
\\
\omega_8^{\ 87}\g_7+\omega_8^{\ 96}\g^8_{\ 96}-e^\Phi G_{0697}\g^{0697}=0
\nonumber
\\
\omega_9^{\ 97}\g_7+\omega_9^{\ 68}\g^9_{\ 68}+e^\Phi G_{0697}\g^{0697}=0
\nonumber
\\
\omega_0^{\ 06}\g_6+e^\Phi H_{0678}\g^{0678}=0
\nonumber
\\
\omega_a^{\ a6}\g_6-e^\Phi H_{0678}\g^{0678}=0
\nonumber
\\
\omega_7^{\ 76}\g_6+\omega_7^{\ 89}\g^7_{\ 89}+e^\Phi H_{0678}\g^{0678}=0
\nonumber
\\
\omega_8^{\ 86}\g_6+\omega_8^{\ 79}\g^8_{\ 79}+e^\Phi H_{0678}\g^{0678}=0
\nonumber
\\
\omega_9^{\ 96}\g_6+\omega_9^{\ 78}\g^9_{\ 78}-e^\Phi H_{0678}\g^{0678}=0
\nonumber
\eea

Some of these equations determine the field strengths
in terms of the spin connections, to be
\bea
\label{fld2}
e^\Phi H_{078a}=-{1\over 2}(\omega_0^{0a}+\omega_7^{7a})
\nonumber
\\
e^\Phi H_{068a}={1\over 2}\omega_6^{a7}
\nonumber
\\
~~~~~~~~~~~~~~~~~~~~~~~~~\qquad e^\Phi H_{067a}=-{1\over 2}
\omega_6^{a8}
\nonumber
\\
e^\Phi H_{0678}=-\omega_0^{06}
\\
e^\Phi G_{069a}={1\over 2}(\omega_0^{0a}-\omega_7^{7a})
\nonumber
\\
e^\Phi G_{09a\a}={1\over 2}\omega_6^{a\a}
\nonumber
\\
e^\Phi G_{069\a}=\omega_0^{0\a}
\nonumber
\eea

The remaining equations yield further constraints on the metric.
\bea
\label{cons2}
\omega_1^{\ 67}\g^1_{\ 67}=\omega_1^{\ 89}\g^1_{\ 89}
\nonumber
\\
\omega_6^{\ 17}\g^6_{\ 17}=-\omega_7^{\ 16}\g^7_{\ 16}=\omega_8^{\ 19}\g^8_{\ 19}
=-\omega_9^{\ 18}\g^9_{\ 18}
\nonumber
\\
\omega_6^{\ 98}\g^6_{\ 98}=(\omega_0^{\ 07}-\omega_6^{\ 67})\g_7
\nonumber
\\
\omega_8^{\ 96}\g^8_{\ 96}=(-\omega_0^{\ 07}-\omega_8^{\ 87})\g_7
\nonumber
\\
\omega_9^{\ 68}\g^9_{\ 68}=(\omega_0^{\ 07}-\omega_9^{\ 97})\g_7
\\
(\omega_7^{\ 76}+\omega_9^{\ 96})\g_6+\omega_7^{\ 89}\g^7_{\ 89}
+\omega_9^{\ 78}\g^9_{\ 78}=0
\nonumber
\\
\omega_7^{\ 89}\g^7_{\ 89}=\omega_8^{\ 79}\g^8_{\ 79}
\nonumber
\\
\omega_7^{\ 89}\g^7_{\ 89}=
(\omega_0^{\ 06}-\omega_8^{\ 86})\g_6
\nonumber
\eea

\subsection{Solution}
We can solve the constraints (\ref{cons1},\ref{cons2}) by the metric ansatz
\bea
e_{0\tilde{0}}=\lambda^{-{1\over 4}}
%\nonumber
%\\
\qquad\qquad
e_{a\tilde{a}}=\lambda^{{1\over 4}}
%\nonumber
%\\
\qquad\qquad
e_{6\tilde{6}}=\lambda^{-{1\over 4}}H^{1\over 2}
\nonumber
\\
e_{7\tilde{7}}=e_{8\tilde{8}}=\lambda^{{1\over 4}}H^{1\over 2}
%\\
\qquad\qquad\qquad
e_{9\tilde{9}}=\lambda^{-{1\over 4}}H^{-{1\over 2}}
%\nonumber
\\
e_{9\tilde{7}}=e_{9\tilde{9}}A_7\qquad\quad e_{9\tilde{8}}=e_{9\tilde{9}}A_8
%\nonumber
%\\
\qquad\qquad
e_{6\tilde{a}}=e_{6\tilde{6}}\phi_a
\nonumber
\eea
with the remaining constraints
\bea
\del_7A_8-\del_8A_7=-\del_6(H\lambda)\qquad\qquad\qquad
\nonumber
\\
\del_6A_8=\del_7 H\qquad\qquad\qquad\qquad\qquad\qquad\quad \del_6A_7=-\del_8  H
\\
\del_6(\phi_aH)=\del_a H\qquad\qquad\qquad\qquad
\nonumber
\eea

Similarly by requiring the variation of the dilatino to vanish,
we find that
the dilaton is given by
\bea
e^\Phi=\lambda^{{1\over 4}}
\eea

We can then show that the field strengths (\ref{fld1},\ref{fld2})
can be obtained from
the gauge fields
\bea
A_{09a}=-{1\over 4}{\phi_a\over \lambda}
\qquad\qquad\qquad\qquad\qquad\qquad A_{069}={1\over 4\lambda}~
\nonumber
\\
 A_{06\a}={A_\a\over 4\lambda}
\qquad\qquad\qquad\qquad\qquad\qquad  A_{0a\a}= {A_\a\phi_a\over 4\lambda}
\\
A_{078}=-{H\over 4}\qquad\qquad\qquad\qquad
\nonumber
\eea

Finally we impose the equations of motion for pointlike sources.
These yield the equation
\bea
\del_a\phi_a=H^{-1}\del_6\lambda+\del_6\left({\phi_a^2\over 2}\right)
\eea

This provides a complete solution for the 
 2-brane on a 2-cycle of a $Z_N$ orbifold.

\subsection{Adding a B-field}

In the solution of the previous section, we were at the point 
in moduli space where the
B-field on the orbifold was zero,
as could be seen from the fact that the
D0-brane charge was zero.
Turning on a B-field on the orbifold point
will add zero-brane charge to the system.

To find a solution with added zero-brane charge, we
lift the solution we have found to 11 dimensions
(thus getting a M2brane on the orbifold.) We then
boost in the 11th direction and dimensionally
reduce, thus obtaining a solution with
D2+D0 charge.

All these steps are straightforward, and we can
directly present the final answer in the next section.
The solutions for other branes on the orbifold can
be obtained by T-duality. 

\section{Branes on orbifolds}
\subsection{Notation}
First of all we summarize the various formulae
that are required in the solutions.

The solutions are expressed in terms of the
functions $\lambda, H, A_7, A_8$. In addition
we will introduce a constant $\b$ and a function $\D$ 
defined as
\bea
\D=cosh^2\b-\lambda^{-1}sinh^2\b
\eea
The constant $\b$ controls the value of the B-field on
the 2-cycle. $\b=0$ corresponds to zero B-field.

The functions satisfy the differential equations
\bea
\del_7A_8-\del_8A_7+\del_6(H\lambda)=Q(x_a,x_6)\d(x_\a)
\nonumber
\\
\del_6A_8=\del_7 H\qquad\qquad\qquad\qquad \quad \del_6A_7=-\del_8  H
\nonumber
\\
\del_6(\phi_aH)=\del_a H\qquad\qquad\qquad
\\
\del_a\phi_a=H^{-1}\del_6\lambda+\del_6\left({\phi_a^2\over 2}\right)\qquad\qquad
\nonumber
\eea

These functions can be expressed in terms of a single function
$\t$ through
\bea
H\phi_a=\del_a\del_6\tau
\\
H=\del_6^2\tau
\\
\l+H\phi_a^2=\del_a^2\tau
\eea

$\t$ then satisfies the differential equation
\bea
\del_a^2\t+\del_6^2\del_a^2\t-(\del_6\del_a\t)^2={1\over \del_6}Q(x_a,x_6)\d(x_\a)
\eea

The function $Q$ parametrizes the brane source; in particular,
it incorporates the effects of brane bending.

We will denote the NSNS 2-form by $B_{\m\n}$, and the RR forms
by $C^{(k)}_{\m_1..\m_k}$.

\subsection{2-brane on $Z_N$ orbifold}
Here $a$ runs from 1 to 5.
The metric is
\bea
ds^2=\lambda^{-{1\over 2}}\D^{-{1\over 2}}(-dt^2+\D dx_a^2)
%+\lambda^{{1\over 2}}\D^{{1\over 2}}dx_a^2
+\lambda^{-{1\over 2}}\D^{{1\over 2}}H(dx^6+\phi_adx^a)^2
+\lambda^{{1\over 2}}\D^{{1\over 2}}H(dx_7^2+dx_8^2)
\nonumber
\\
+\lambda^{-{1\over 2}}\D^{{1\over 2}}H^{-1}(dx_9+A_7dx^7+A_8dx_8)^2
\eea
%\bea
%e_{0\tilde{0}}=\lambda^{-{1\over 4}}\D^{-{1\over 4}}
%\nonumber
%\\
%e_{a\tilde{a}}=\lambda^{{1\over 4}}\D^{{1\over 4}}
%\nonumber
%\\
%e_{6\tilde{6}}=\lambda^{-{1\over 4}}H^{1\over 2}\D^{{1\over 4}}
%\nonumber
%\\
%e_{7\tilde{7}}=e_{8\tilde{8}}=\lambda^{{1\over 4}}H^{1\over 2}\D^{{1\over 4}}
%\\
%e_{9\tilde{9}}=\lambda^{-{1\over 4}}H^{-{1\over 2}}\D^{{1\over 4}}
%\nonumber
%\\
%e_{9\tilde{7}}=e_{9\tilde{9}}A_7\quad e_{9\tilde{8}}=e_{9\tilde{9}}A_8
%\nonumber
%\\
%e_{6\tilde{a}}=e_{6\tilde{6}}\phi_a
%\nonumber
%\eea

The dilaton is 
\bea
e^\Phi=\lambda^{{1\over 4}}\D^{{3\over 4}}
\eea

The gauge fields are
\bea
C^{(1)}_0=sinh\b cosh\b{(1-\l)\over \D\l}~~~~~~~~~~~~~~~~~~~~~
\nonumber
\\
C^{(3)}_{09a}=-{1\over 4}{\phi_a\over \lambda}cosh\b
~~~~~~~~~~~~~~~~~~~~~~~~~~\quad C^{(3)}_{069}={1\over 4\lambda}cosh\b
\nonumber
\\
 C^{(3)}_{06\a}={A_\a\over 4\lambda}cosh\b
~~~~~~~~~~~~~~~~~~~~~~~~~\quad  C^{(3)}_{0a\a}= {A_\a\phi_a\over 4\lambda}cosh\b
\nonumber
\\
C^{(3)}_{078}=-{H\over 4}cosh\b~~~~~~~~~~~~~~~~~~~~~~~~~~
\\
B_{9a}=-{3\over 8}{\phi_a\over \lambda}sinh\b
~~~~~~~~~~~~~~~~~~~~~~~~~~\quad B_{69}={3\over 8\lambda}sinh\b
\nonumber
\\
 B_{6\a}={3A_\a\over 8\lambda}sinh\b
~~~~~~~~~~~~~~~~~~~~~~~~~~\quad  B_{a\a}= {3A_\a\phi_a\over 8\lambda}sinh\b
\nonumber
\\
B_{78}=-{3H\over 8}sinh\b~~~~~~~~~~~~~~~~~~~~~~~~~~
\nonumber
\eea

\subsection{3-brane on $Z_N$ orbifold}

Here $i$ runs over $0,1$, $a$ runs over $2,3,4,5$.
The metric is
\bea
ds^2=\lambda^{-{1\over 2}}\D^{-{1\over 2}}(-dt^2+dx_1^2+\D dx_a^2)
%+\lambda^{{1\over 2}}\D^{{1\over 2}}dx_a^2
+\lambda^{-{1\over 2}}\D^{{1\over 2}}H(dx^6+\phi_adx^a)^2
+\lambda^{{1\over 2}}\D^{{1\over 2}}H(dx_7^2+dx_8^2)
\nonumber
\\
+\lambda^{-{1\over 2}}\D^{{1\over 2}}H^{-1}(dx_9+A_7dx^7+A_8dx_8)^2
\eea

%\bea
%e_{i\tilde{i}}=\lambda^{-{1\over 4}}\D^{-{1\over 4}}
%\nonumber
%\\
%e_{a\tilde{a}}=\lambda^{{1\over 4}}\D^{{1\over 4}}
%\nonumber
%\\
%e_{6\tilde{6}}=\lambda^{-{1\over 4}}H^{1\over 2}\D^{{1\over 4}}
%\nonumber
%\\
%e_{7\tilde{7}}=e_{8\tilde{8}}=\lambda^{{1\over 4}}H^{1\over 2}\D^{{1\over 4}}
%\\
%e_{9\tilde{9}}=\lambda^{-{1\over 4}}H^{-{1\over 2}}\D^{{1\over 4}}
%\nonumber
%\\
%e_{9\tilde{7}}=e_{9\tilde{9}}A_7\quad e_{9\tilde{8}}=e_{9\tilde{9}}A_8
%\nonumber
%\\
%e_{6\tilde{a}}=e_{6\tilde{6}}\phi_a
%\nonumber
%\eea

The dilaton is
\bea
e^\Phi=\D^{{1\over 2}}
\eea

The gauge fields are
\bea
C^{(2)}_{01}=sinh\b cosh\b{(1-\l)\over \D\l}~~~~~~~~~~~~~~~~~~~~~~~~~
~~~~~~~
\nonumber
\\
C^{(4)}_{abc7}={1\over 4}\e_{abcd}(\phi_dA_7+\del_8\del_d\tau)
~~~~~~~~~~~~~~~~~~~~~~~
C^{(4)}_{abc8}={1\over 4}\e_{abcd}(\phi_dA_8-\del_7\del_d\tau)
\nonumber
\\
C^{(4)}_{abc9}={1\over 4}\e_{abcd}\phi_d
~~~~~~~~~~~~~~~~~~~~~~~~~~~~~~~~~~~~~~~~
%A_{09a}=-{1\over 4}{\phi_a\over \lambda}cosh\b
%\quad A_{069}={1\over 4\lambda}cosh\b
%\nonumber
%\\
 %A_{06\a}={A_\a\over 4\lambda}cosh\b
%\quad  A_{0a\a}= {A_\a\phi_a\over 4\lambda}cosh\b
%\nonumber
%\\
%A_{078}=-{H\over 4}cosh\b
\nonumber
\\
B_{9a}=-{3\over 8}{\phi_a\over \lambda}sinh\b
~~~~~~~~~~~~~~~~~~~~~~~~~~\quad B_{69}={3\over 8\lambda}sinh\b
~~~~~~~~~~~~~~~~~
\nonumber
\\
 B_{6\a}={3A_\a\over 8\lambda}sinh\b
~~~~~~~~~~~~~~~~~~~~~~~~~~\quad  B_{a\a}= {3A_\a\phi_a\over 8\lambda}sinh\b
\nonumber
~~~~~~~~~~~~
\\
B_{78}=-{3H\over 8}sinh\b~~~~~~~~~~~~~~~~~~~~~~~~~~
~~~~~~~~~~~~~~~~~
\nonumber
\eea

\subsection{4-brane on $Z_N$ orbifold}

Here $i$ runs over $1,2$, $a$ runs over $3,4,5$.
The metric is
\bea
ds^2=\lambda^{-{1\over 2}}\D^{-{1\over 2}}(-dt^2+dx_i^2+\D dx_a^2)
%+\lambda^{{1\over 2}}\D^{{1\over 2}}dx_a^2
+\lambda^{-{1\over 2}}\D^{{1\over 2}}H(dx^6+\phi_adx^a)^2
+\lambda^{{1\over 2}}\D^{{1\over 2}}H(dx_7^2+dx_8^2)
\nonumber
\\
+\lambda^{-{1\over 2}}\D^{{1\over 2}}H^{-1}(dx_9+A_7dx^7+A_8dx_8)^2
\eea

%Here $i$ runs over $0,1$, $a$ runs over $2,3,4,5$.
%\bea
%e_{i\tilde{i}}=\lambda^{-{1\over 4}}\D^{-{1\over 4}}
%\nonumber
%\\
%\nonumber
%%e_{a\tilde{a}}=\lambda^{{1\over 4}}\D^{{1\over 4}}
%\nonumber
%\\
%e_{6\tilde{6}}=\lambda^{-{1\over 4}}H^{1\over 2}\D^{{1\over 4}}
%\nonumber
%\\
%e_{7\tilde{7}}=e_{8\tilde{8}}=\lambda^{{1\over 4}}H^{1\over 2}\D^{{1\over 4}}
%\\
%e_{9\tilde{9}}=\lambda^{-{1\over 4}}H^{-{1\over 2}}\D^{{1\over 4}}
%\nonumber
%\\
%e_{9\tilde{7}}=e_{9\tilde{9}}A_7\quad e_{9\tilde{8}}=e_{9\tilde{9}}A_8
%\nonumber
%\\
%e_{6\tilde{a}}=e_{6\tilde{6}}\phi_a
%\nonumber
%%\eea

The dilaton is
\bea
e^\Phi=\lambda^{-{1\over 4}}\D^{{1\over 4}}
\eea

The gauge fields are
\bea
C^{(3)}_{012}=sinh\b cosh\b{(1-\l)\over \D\l}~~~~~~~~~~~~~~~~~~~~~~~~~
~~~~~~~
\nonumber
\\
C^{(3)}_{ab7}={1\over 4}\e_{abc}(\phi_cA_7+\del_8\del_c\tau)
~~~~~~~~~~~~~~~~~~~~~~~
C^{(3)}_{ab8}={1\over 4}\e_{abc}(\phi_cA_8-\del_7\del_c\tau)
\nonumber
\\
C^{(3)}_{ab9}={1\over 4}\e_{abc}\phi_c
~~~~~~~~~~~~~~~~~~~~~~~~~~~~~~~~~~~~~~~~
\nonumber
\\
B_{9a}=-{3\over 8}{\phi_a\over \lambda}sinh\b
~~~~~~~~~~~~~~~~~~~~~~~~~~\quad B_{69}={3\over 8\lambda}sinh\b
~~~~~~~~~~~~~~~~~
\nonumber
\\
 B_{6\a}={3A_\a\over 8\lambda}sinh\b
~~~~~~~~~~~~~~~~~~~~~~~~~~\quad  B_{a\a}= {3A_\a\phi_a\over 8\lambda}sinh\b
\nonumber
~~~~~~~~~~~~
\\
B_{78}=-{3H\over 8}sinh\b~~~~~~~~~~~~~~~~~~~~~~~~~~
~~~~~~~~~~~~~~~~~
\nonumber
\eea

\subsection{5-brane on $Z_N$ orbifold}

Here $i$ runs over $1,2,3$, $a$ runs over $4,5$.
The metric is
\bea
ds^2=\lambda^{-{1\over 2}}\D^{-{1\over 2}}(-dt^2+dx_i^2+\D dx_a^2)
%+\lambda^{{1\over 2}}\D^{{1\over 2}}dx_a^2
+\lambda^{-{1\over 2}}\D^{{1\over 2}}H(dx^6+\phi_adx^a)^2
+\lambda^{{1\over 2}}\D^{{1\over 2}}H(dx_7^2+dx_8^2)
\nonumber
\\
+\lambda^{-{1\over 2}}\D^{{1\over 2}}H^{-1}(dx_9+A_7dx^7+A_8dx_8)^2
\eea

The dilaton is
\bea
e^\Phi=\lambda^{-{1\over 2}}
\eea

The gauge fields are
\bea
C^{(4)}_{0123}=sinh\b cosh\b{(1-\l)\over \D\l}~~~~~~~~~~~~~~~~~~~~~~~~~
~~~~~~~
\nonumber
\\
C^{(2)}_{a7}={1\over 4}\e_{ab}(\phi_bA_7+\del_8\del_b\tau)
~~~~~~~~~~~~~~~~~~~~~~~
C^{(2)}_{a8}={1\over 4}\e_{ab}(\phi_bA_8-\del_7\del_b\tau)
\nonumber
\\
C^{(2)}_{a9}={1\over 4}\e_{ab}\phi_b
~~~~~~~~~~~~~~~~~~~~~~~~~~~~~~~~~~~~~~~~
\nonumber
\\
B_{9a}=-{3\over 8}{\phi_a\over \lambda}sinh\b
~~~~~~~~~~~~~~~~~~~~~~~~~~\quad B_{69}={3\over 8\lambda}sinh\b
~~~~~~~~~~~~~~~~~
\nonumber
\\
 B_{6\a}={3A_\a\over 8\lambda}sinh\b
~~~~~~~~~~~~~~~~~~~~~~~~~~\quad  B_{a\a}= {3A_\a\phi_a\over 8\lambda}sinh\b
\nonumber
~~~~~~~~~~~~
\\
B_{78}=-{3H\over 8}sinh\b~~~~~~~~~~~~~~~~~~~~~~~~~~
~~~~~~~~~~~~~~~~~
\nonumber
\eea

\section{Comments on the duality}

The last subsection above describes the supergravity
solution for 5-branes wrapped on an orbifold 2-cycle.
By taking the near-horizon limit, we obtain the
dual to ${\cal N}=2$ gauge theory in 3+1 dimensions.

The orbifold point is defined by the point in the moduli
space where $B={1\over N}$ for a $Z_N$ orbifold.
At this point in moduli space, a single 2-brane
wrapped on the orbifold gets an induced
0-brane charge equal to ${1\over N}$. This
can be used to fixed the value of $\beta$
at the orbifold point.

The calculation is straightforward, given the supergravity
solution. We find that (we set $\a'=1$)
\bea
sinh \beta= {1\over 4\pi Na}
\eea
(Recall that $a$ is the separation between the
centers in the orbifold metric (\ref{defa}).)
This value of $\beta$ is the perturbative orbifold point. Other
values of $\beta$ correspond to changing the B-field
on the 2-cycle.

One should also ask where the supergravity solution we have
found is valid. For this, we need 
the string coupling to be small. 
Hence
\bea
e^\Phi=g_0\lambda^{-{1\over 2}}\ll 1
\eea
This means that the supergravity solution cannot
be trusted near the branes. This corresponds to the
fact that that the gauge theory is trivial in the
infrared, and cannot therefore have a
simple description in supergravity.

Also, we need the curvature to be small. We cannot
find the exact regime where this applies, since 
we have not found an analytic solution.
However, we know that this must break down
far away from the branes, since asymptotically, the
geometry is similar to that of a five-brane.
This breakdown corresponds to the fact that the
gauge theory is asymptotically free, and again there cannot
be a good supergravity description.

Finally, we can see that in the weak coupling
region, the gauge coupling has a logarithmic
falloff with scale, as is expected from the gauge
theory correspondence. This can be seen from
the fact that the gauge coupling asymptoticaly satisfies
the Laplace equation in 2 space dimensions, which
has a logarithmic solution. This analysis is the same
as that performed in the geometric engineering
analyses of field theories \cite{Katz:1997th,Katz:1997fh,Katz:1998eq}. 

\section{Conclusion}

In summary, we have constructed supergravity duals
for ${\cal N}=2$ gauge theories. This method can
be extended to all theories which can be
obtained on branes through geometric engineering.
This is a very wide class of theories.

We have left the detailed analysis of the correspondence
to future work. It
would be interesting to see what information can be
obtained about the gauge theories by this duality.
In particular, it should be possible to rederive the
results of \cite{Seiberg:1994rs,Douglas:1995nw}.
It should also be possible to analyze the behaviour of
solitons in these theories by analyzing
branes ending on other branes.

\section{Acknowledgements}

This research was supported in part by DOE grant DE-FG02-96ER40559.


\begin{thebibliography}   {99}

\bibitem{Maldacena:1998re}
J.~Maldacena,
%``The large N limit of superconformal field theories and supergravity,''
Adv.\ Theor.\ Math.\ Phys.\  {\bf 2}, 231 (1998)
[hep-th/9711200].

%PolchStrass

%\cite{Polchinski:2000uf}:
\bibitem{Polchinski:2000uf}
J.~Polchinski and M.~J.~Strassler,
%``The string dual of a confining four-dimensional gauge theory,''
hep-th/0003136.


% KlebanTseyt; KlebStras;

\bibitem{Klebanov:2000nc}
I.~R.~Klebanov and A.~A.~Tseytlin,
%``Gravity duals of supersymmetric SU(N) x SU(N+M) gauge theories,''
Nucl.\ Phys.\  {\bf B578}, 123 (2000)
[hep-th/0002159].
%%CITATION = HEP-TH 0002159;%%

%\cite{Klebanov:2000hb}:
\bibitem{Klebanov:2000hb}
I.~R.~Klebanov and M.~J.~Strassler,
%``Supergravity and a confining gauge theory: Duality cascades and  (chi)SB-resolution of naked singularities,''
JHEP {\bf 0008}, 052 (2000)
[hep-th/0007191].
%%CITATION = HEP-TH 0007191;%%


% Acharya

%\cite{Acharya:2000gb}:
\bibitem{Acharya:2000gb}
B.~S.~Acharya,
%``On realising N = 1 super Yang-Mills in M theory,''
hep-th/0011089.
%%CITATION = HEP-TH 0011089;%%

%\cite{Maldacena:2000yy}:
\bibitem{Maldacena:2000yy}
J.~M.~Maldacena and C.~Nunez,
%``Towards the large n limit of pure N = 1 super Yang Mills,''
hep-th/0008001.
%%CITATION = HEP-TH 0008001;%%

\bibitem{MV}
M.~Atiyah, J.~M.~Maldacena and C.~Vafa,
hep-th/0011256.
%Brinne

%\cite{Brinne:2000fh}:
\bibitem{Brinne:2000fh}
B.~Brinne, A.~Fayyazuddin, S.~Mukhopadhyay and D.~J.~Smith,
%``Supergravity M5-branes wrapped on Riemann surfaces and their QFT duals,''
hep-th/0009047.
%%CITATION = HEP-TH 0009047;%%

%\cite{Witten:1998jd}:
\bibitem{Witten:1998jd}
E.~Witten,
%``Branes and the dynamics of QCD,''
%in {\it NONE}
Nucl.\ Phys.\ Proc.\ Suppl.\  {\bf 68}, 216 (1998).
%%CITATION = NUPHZ,68,216;%%

%\cite{Rajaraman:2000ws}:
\bibitem{Rajaraman:2000ws}
A.~Rajaraman,
%``Supergravity solutions for localised brane intersections,''
hep-th/0007241.
%%CITATION = HEP-TH 0007241;%%


%DogMoore

%\cite{Douglas:1996sw}:
\bibitem{Douglas:1996sw}
M.~R.~Douglas and G.~Moore,
%``D-branes, Quivers, and ALE Instantons,''
hep-th/9603167.
%%CITATION = HEP-TH 9603167;%%

%MyersJohns
%\cite{Johnson:1997py}:
\bibitem{Johnson:1997py}
C.~V.~Johnson and R.~C.~Myers,
%``Aspects of type IIB theory on ALE spaces,''
Phys.\ Rev.\  {\bf D55}, 6382 (1997)
[hep-th/9610140].
%%CITATION = HEP-TH 9610140;%%

%\cite{Aspinwall:1995zi}:
\bibitem{Aspinwall:1995zi}
P.~S.~Aspinwall,
%``Enhanced gauge symmetries and K3 surfaces,''
Phys.\ Lett.\  {\bf B357}, 329 (1995)
[hep-th/9507012].
%%CITATION = HEP-TH 9507012;%%


%\cite{Polchinski:2000mx}:
\bibitem{Polchinski:2000mx}
J.~Polchinski,
%``N = 2 gauge-gravity duals,''
hep-th/0011193.
%%CITATION = HEP-TH 0011193;%%

%VafaKatz

%\cite{Katz:1997th}:
\bibitem{Katz:1997th}
S.~Katz and C.~Vafa,
%``Geometric engineering of N = 1 quantum field theories,''
Nucl.\ Phys.\  {\bf B497}, 196 (1997)
[hep-th/9611090].
%%CITATION = HEP-TH 9611090;%%

%\cite{Katz:1997fh}:
\bibitem{Katz:1997fh}
S.~Katz, A.~Klemm and C.~Vafa,
%``Geometric engineering of quantum field theories,''
Nucl.\ Phys.\  {\bf B497}, 173 (1997)
[hep-th/9609239].
%%CITATION = HEP-TH 9609239;%%


%\cite{Katz:1998eq}:
\bibitem{Katz:1998eq}
S.~Katz, P.~Mayr and C.~Vafa,
%``Mirror symmetry and exact solution of 4D N = 2 gauge theories. I,''
Adv.\ Theor.\ Math.\ Phys.\  {\bf 1}, 53 (1998)
[hep-th/9706110].
%%CITATION = HEP-TH 9706110;%%

%Aspinwall



%SeibergWitten
%\cite{Seiberg:1994rs}:
\bibitem{Seiberg:1994rs}
N.~Seiberg and E.~Witten,
%``Electric - magnetic duality, monopole condensation, and confinement in N=2 supersymmetric Yang-Mills theory,''
Nucl.\ Phys.\  {\bf B426}, 19 (1994)
[hep-th/9407087].
%%CITATION = HEP-TH 9407087;%%




%\cite{Douglas:1995nw}:
\bibitem{Douglas:1995nw}
M.~R.~Douglas and S.~H.~Shenker,
%``Dynamics of SU(N) supersymmetric gauge theory,''
Nucl.\ Phys.\  {\bf B447}, 271 (1995)
[hep-th/9503163].
%%CITATION = HEP-TH 9503163;%%

\end{thebibliography}
\end{document}